\documentclass[10pt]{extarticle}
\usepackage[bordercolor=white,backgroundcolor=gray!30,linecolor=black,colorinlistoftodos, textsize=10pt]{todonotes}

\usepackage{amssymb}
\usepackage{textcomp}
\usepackage{amsmath}
\usepackage[textstyle,amssymb]{SIunits}
\usepackage{graphicx}
\usepackage[twocolumn,textwidth=175mm,columnsep=6mm,top=20mm,bottom=25mm]{geometry}
\usepackage{helvet}
\usepackage{titlesec}
\titleformat*{\section}{\bfseries\sffamily}
\titlespacing{\section}{0pt}{*4}{*0}
\titleformat{\subsection}[runin]{\normalfont\bfseries}{\thesubsection.}{3pt}{}
\usepackage{setspace}
\usepackage[breaklinks=true]{hyperref} 
\usepackage[font={footnotesize,sf},labelfont={sf,bf},labelsep=endash,justification=raggedright]{caption}

\usepackage{natbib}
\usepackage[labelfont=bf,justification=justified]{caption}
\usepackage{stfloats}
\usepackage{xr}
\usepackage{lineno}
\begin{document}

\twocolumn[\begin{@twocolumnfalse}
	{\huge\sf \noindent\textbf{Dissipative Kerr solitons in semiconductor ring lasers}}
	\vspace{0.5cm}
	
	{\sf\large\noindent \textbf {Bo Meng$^{1*}$, Matthew Singleton$^1$, Johannes Hillbrand$^1$, Martin Franckié$^1$, Mattias Beck$^1$ and J\'er\^ome Faist$^{1*}$}}
		\vspace{0.5cm}
		
		{\sf \noindent\textbf{$^1$Institute for Quantum Electronics, ETH Zurich, 8093 Zurich, Switzerland\newline}
				$^*$e-mail: {bmeng@phys.ethz.ch};
				$^*$e-mail: {jerome.faist@phys.ethz.ch}}
		\vspace{0.5cm}


{\sf \small \textbf{
		\noindent Dissipative Kerr solitons are self-organized optical waves arising from the interplay between Kerr effect and dispersion. They can form spontaneously in nonlinear microresonators pumped with an external continuous-wave laser, which provides the parametric gain for the proliferation of an ultrastable frequency comb. These miniaturized and battery driven microcombs have become a disruptive technology for precision metrology, broadband telecommunication and ultrafast optical ranging. In this work, we report on the first experimental observation of dissipative Kerr solitons generated in a ring cavity with a fast semiconductor gain medium. The moderate quality factor of the ring cavity is compensated by the giant resonant Kerr nonlinearity of a quantum cascade laser, which is more than a million times larger than in Si$_3$N$_4$. By engineering the dispersion of the cavity, we observe the formation of bright dissipative Kerr solitons in the mid-infrared range. Soliton formation appears after an abrupt symmetry breaking between the two lasing directions of the ring cavity. The pump field of the soliton is generated by direct electrical driving and closely resembles the soliton Cherenkov radiation observed in passive microcombs. Two independent techniques shed light on the waveform and coherence of the solitons and confirm a pulse width of approximately 3~ps. Our results extend the spectral range of soliton microcombs to mid-infrared wavelengths and will lead to integrated, battery driven and turnkey spectrometers in the molecular fingerprint region.
	}
}
\vspace{0.5cm}
\end{@twocolumnfalse}]

\section*{Introduction}
\noindent A dissipative Kerr soliton is a self-organized optical wave defined by its ability to propagate through a dispersive, nonlinear and lossy medium while preserving its shape and amplitude\cite{grelu2012dissipative,kippenberg2018dissipative}. The spreading of the wave packet due to dispersion is balanced by self-phase modulation induced by the Kerr effect. Likewise, its amplitude is maintained by parametric gain provided by an external pump laser. Dissipative Kerr solitons were first demonstrated in active and passive fiber resonators\cite{kivshar2003optical,grelu2012dissipative,leo2010temporal} and have recently witnessed a surge of attention sparked by the development of integrated microresonators with ultrahigh quality factor (Q-factor)\cite{herr2014temporal,matsko2011mode}. These microcombs have already proven to be a disruptive technology for coherent telecommunication\cite{marin2017microresonator}, ultrafast optical ranging \cite{trocha2018ultrafast}, precision metrology\cite{picque2019frequency,suh2016microresonator} and frequency synthesis\cite{spencer2018optical,weng2020frequency}. The rapid progress of photonic integration during the recent years has enabled low-loss microresonators based on silicon nitride\cite{liu2021high} and silicon dioxide\cite{wu2020greater} with Q-factors routinely exceeding 10$^7$. This advancement led to a dramatic reduction of the threshold pump power for soliton generation, which scales inversely with the square of the Q-factor, and thus enabled a new generation of miniaturized and battery driven frequency combs \cite{stern2018battery,Shen2020integrated} compatible with silicon mass production\cite{liu2021high}. More recently, III-V semiconductors such as AlGaAs\cite{chang2020ultra,moille2020dissipative} and GaP\cite{wilson2019integrated} are emerging as promising candidates for microcomb generation. Although the Q-factor achieved with these materials is not yet on par with Si$_3$N$_4$ technology, their Kerr nonlinearity is 1 to 2 orders of magnitude larger. As a result, ultra-efficient Kerr combs with merely 36\,\textmu W threshold power were observed in AlGaAs microresonators.

The examples above showcase the trade-off between the ultrahigh Q-factors enabled by Si$_3$N$_4$ and SiO$_2$ microresonators and the large Kerr nonlinearity provided by III-V materials. In this letter, we leverage on the giant nonlinearity of intersubband transitions in  quantum wells to enable mid-infrared (mid-IR) soliton microcombs by direct electrical pumping. Instead of a passive microresonator with external pumping, we employ a semiconductor laser embedded in a ring cavity. The semiconductor gain medium simultaneously serves two purposes. Firstly, the laser transition gives rise to a giant resonant Kerr nonlinearity, which is many orders of magnitude larger than in bulk materials. Secondly, the gain medium allows to generate the pump field required for dissipative Kerr solitons directly in the ring cavity without the need for external pumping. This property is especially advantageous for resonators with only a moderate Q-factor. In this case, the threshold intracavity power for parametric oscillation scales linearly with the cavity losses. Hence, the stringent requirement of a high Q-factor is significantly reduced.

The quantum cascade laser\cite{faist1994quantum} (QCL) constitutes an ideal technology for generating Kerr frequency combs. During the past two decades, QCLs have matured to become the workhorse of mid-IR spectroscopy featuring Watt-level output power\cite{jouy2017dual} and broad spectral coverage\cite{hugi2009external}. Most importantly, QCLs are endowed with a giant and ultrafast Kerr nonlinearity\cite{friedli2013four,opacak2021frequency} on the order of 10$^{-14}$ to 10$^{-13}$\,m$^2$/W\cite{friedli2013four,cai2015investigation} - up to 6 orders of magnitude larger than in Si$_3$N$_4$. This nonlinearity originates from the gain medium and is resonantly enhanced by the ultrafast gain dynamics and the large dipole moment of the laser levels. 
Similar to soliton microcombs, four-wave-mixing leads to the formation of coherent frequency combs in Fabry-P\'erot QCLs \cite{hugi2012mid}. In contrast, these QCL frequency combs are characterized by a quasi-continuous waveform combined with a linear frequency chirp~\cite{singleton2018evidence}. This behavior is a consequence of the instantaneous gain grating arising from the two counter-propagating waves in the Fabry-P\'erot cavity~\cite{khurgin2014coherent}. These frequency modulated combs were observed in a wide variety of semiconductor lasers\cite{schwarz2019monolithic,sterczewski2020frequency,kriso2021signatures} and can be seen as a phase soliton\cite{burghoff2020unraveling}.

\begin{figure*}[t!]
    \centering
    \includegraphics[width=\textwidth]{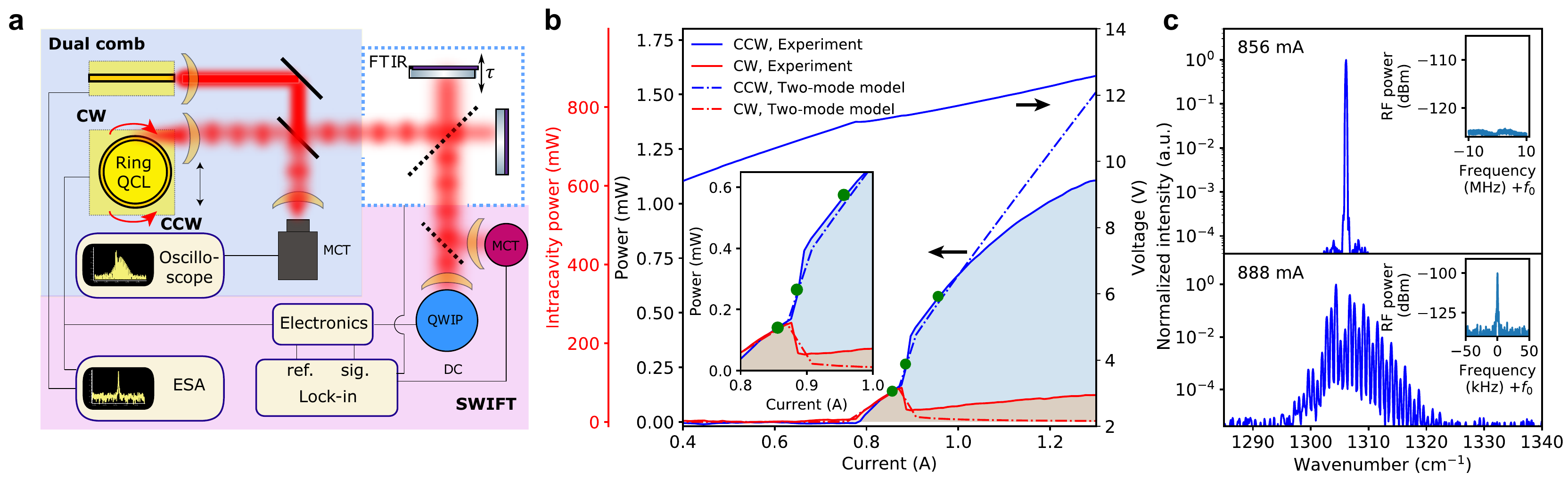}
    \caption{\textbf{Dissipative Kerr temporal soliton in a ring quantum cascade lasers (QCL).} \textbf{a}, Experimental setup for soliton characterization. The spectra and phases of the QCL are measured by coherent beatnote spectroscopy (SWIFTS) and dual-comb spectroscopy. \textbf{b}, Light-current-voltage (L-I-V) curve of a ring QCL at -20~$^{\circ}$C. Symmetry breaking between the clockwise and counterclockwise lasing directions of a ring QCL is observed around 880~mA. The green dots indicate the currents for which the optical spectra are shown. Inset: zoom-in view of the L-I curves for both lasing directions around the symmetry breaking point.\textbf{c}, Optical spectra measured at currents below and above the symmetry breaking point, respectively. Insets: radio-frequency (RF) beatnotes extracted from the device via a bias-tee around $f_0 \approx$ 23.87 GHz. The resolution bandwidth of the spectrum analyser is 200 Hz.}
    \label{fig:fig1}
\end{figure*}

More recently, significant efforts are aimed at generating frequency combs in QCLs with ring cavities~\cite{meng2020mid,piccardo2020frequency}, where the gain grating is suppressed and the existence of long-sought dissipative Kerr solitons was predicted theoretically\cite{columbo2021unifying}. Although previous work reported on spectra which can be approximately fitted with a hyperbolic secant (sech$^2$) envelope~\cite{meng2020mid,piccardo2020frequency}, the observed comb states were not consistent with solitons. Therefore, soliton generation from QCLs remains elusive until now. Here, We unambiguously show that bright soliton generation can indeed take place in ring QCLs under two essential conditions. Firstly, minimal optical backscattering is required to allow unidirectional operation through spontaneous symmetry breaking~\cite{Angelo19912spatiotemporal}. Secondly, the ring cavity is engineered to provide anomalous dispersion, which enables the subtle balance between Kerr effect and dispersion characteristic for solitons.
At the same time, the intracavity power exceeds the threshold for parametric oscillation, which is enabled by a large Kerr nonlinearity and low optical loss in our device.


\section*{Soliton formation}

\begin{figure}[t!]
    \centering\includegraphics[width=0.45\textwidth]{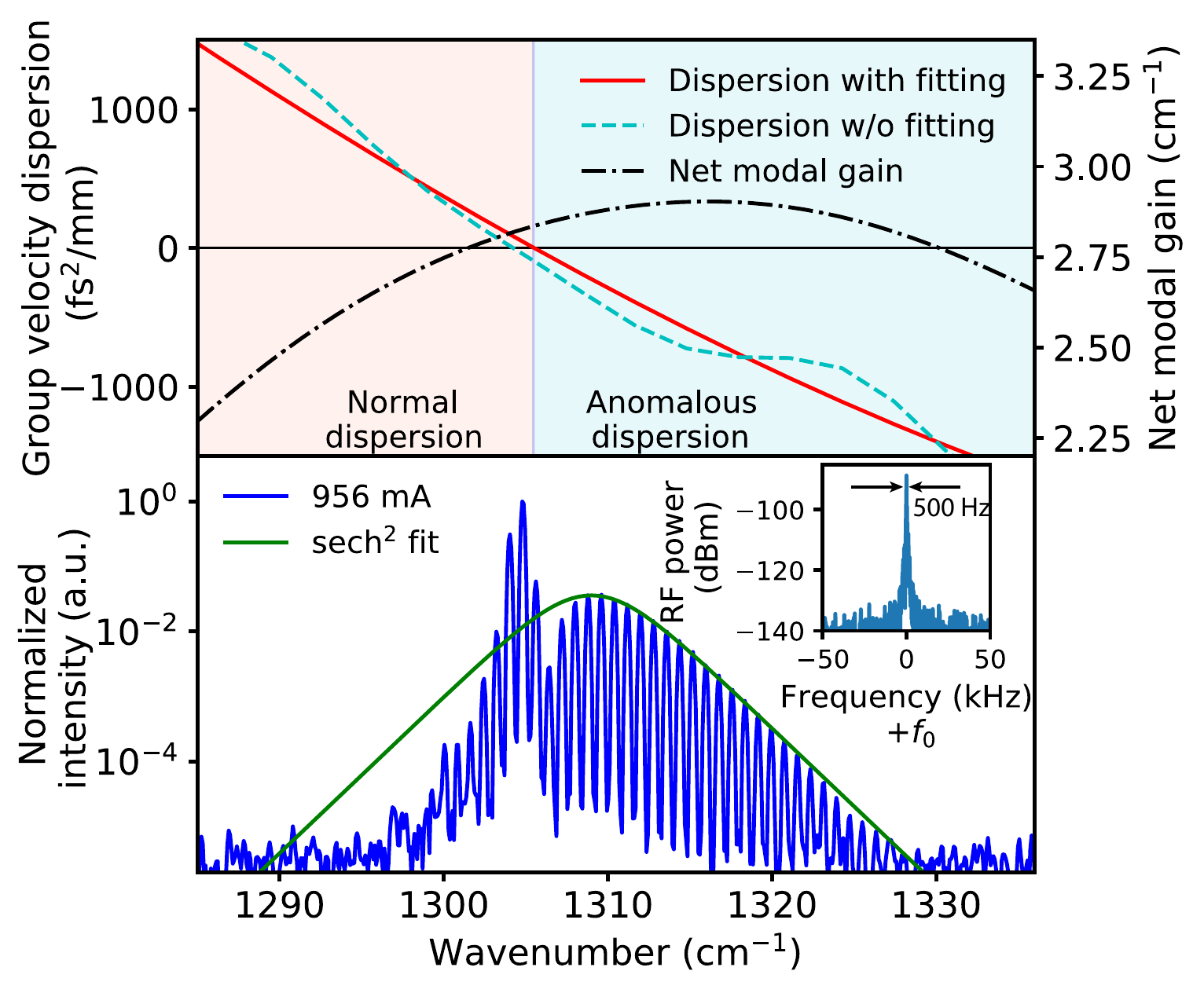}
    \caption{\textbf{Top plot}, Dispersion and gain characteristics of the soliton device. The measurement was carried out by analysing the subthreshold luminescence of a half-ring structure with the same cavity dimensions as the soliton device \cite{hofstetter1999measurement} . The red solid line (cyan dashed line) is calculated by taking the second order derivative of phase with respect to the circular frequency, with (without) the phase fitted with a 3rd-order polynomial. 
    \textbf{Bottom plot}, Optical spectrum of the soliton state at a current of 956 mA and the fit of the state with a spectral sech$^2$ envelope. Inset: RF beatnote of the soliton state around $f_0 =$23.86   GHz. The resolution bandwidth of the spectrum analyser is 200 Hz. 
    }
    \label{fig:fig2}
\end{figure}

In order to ensure unidirectional lasing, the QCL ring cavities were fabricated into buried heterostructure waveguides\cite{beck2001continuous} providing extremely smooth sidewalls and very low optical losses (see Methods). Backscattering is further minimized by the low refractive index contrast between the active region and the InP cladding.
Moreover, the top cladding is capped with a highly doped InP layer. The resulting plasmon mode introduces the anomalous group velocity dispersion\cite{bidaux2017plasmon} (GVD) essential for soliton formation\cite{herr2014temporal}.
As shown schematically in Fig.~\ref{fig:fig1}a, the light emitted tangentially along the ring by the bending losses is collected by a lens. In this way, the clockwise (CW) and counterclockwise (CCW) lasing directions can be selected by merely moving the lens across the cleaved facet.This technique for collecting the power minimizes the perturbations to the ring and thus backscattering.
The collected and intracavity powers in both directions (CW and CCW) are displayed as a function of injected current in Fig.~\ref{fig:fig1}b. Just above laser threshold at I= 780~mA, the device first emits equal powers in both lasing directions and remains monochromatic (Fig.~\ref{fig:fig1}c). As the current is further increased, an abrupt symmetry breaking is observed at 873~mA and power is transferred to the CCW direction at the cost of the CW direction. This spontaneous symmetry breaking goes along with a transition to multimode operation exhibiting an extremely narrow beatnote (Fig.~\ref{fig:fig2}c bottom).
Symmetry breaking between the two counter-rotating modes is a fundamental feature of ring lasers that arises because 
the self-gain saturation is only half as large as cross-gain saturation~\cite{Angelo19912spatiotemporal}. This tendency for unidirectional operation is competing with backscattering having the opposite effect. The basic physics of this effect are fully captured by a pair of differential equations~\cite{Angelo19912spatiotemporal,sorel2003operating}, whose results are shown in Fig.~\ref{fig:fig1}b in dashed lines (see Supplementary Information). The symmetry breaking point allows us to estimate roughly the strength of the backscattering coefficient to $\alpha_{back} = 0.01 $ cm$^{-1}$. 
\
The state of the ring QCL changes drastically above the symmetry breaking point and assumes a shape, which is well known from soliton microcombs (Fig. \ref{fig:fig2} bottom). The spectrum consists of a few strong modes superimposed on a sech$^2$-shaped lobe and exhibits a remarkably strong and narrow beatnote. As strictly required for solitons, the sech$^2$ lobe is located in a spectral region where the ring cavity provides anomalous dispersion (Fig.~\ref{fig:fig2} top). The peak of the spectrum arises at the zero-crossing of the dispersion and can be interpreted qualitatively as a dispersive wave (or analogously, Cherenkov radiation)~\cite{akhmediev1995cherenkov,brasch2016photonic}. The latter effectively replaces the external optical pump of Kerr microresonators.

While the spectrum shown in Fig. \ref{fig:fig2} strongly indicates the presence of solitons, a rigorous proof for dissipative Kerr solitons requires a measurement of the spectral phases emitted by the QCL. This challenging task - especially in the mid-infrared range - has been facilitated recently thanks to the development of phase-sensitive heterodyne techniques. 
In this work, we employ shifted wave interference Fourier transform spectroscopy\cite{burghoff2015evaluating} (SWIFTS) and dual-comb spectroscopy\cite{cappelli2019retrieval} to independently confirm the generation of temporal solitons in the ring QCL (see Methods).
SWIFTS relies on a high-speed quantum well infrared photodetector (QWIP) to measure the laser beatnote. The individual intermode beatings, which carry the phase information, are isolated using an FTIR spectrometer acting as narrowband spectral filter. The excellent accuracy of SWIFTS for pulse characterisation has been already proven in mode-locked Ti-sapphire\cite{han2020sensitivity}, quantum dot\cite{hillbrand2020inphase} and quantum cascade lasers\cite{hillbrand2020mode}. On the other hand, dual-comb spectroscopy relies on a second reference comb with a slightly different mode spacing. The spectrum of the ring comb is downconverted from the optical to the RF domain by combining it with the reference comb on a fast photodetector. This gives rise to a multiheterodyne beating between the comb lines, whose Fourier phases enable the extraction of the spectral phases of the ring QCL. 

\begin{figure*}[ht]
	\centering
	\includegraphics[width=0.98\linewidth]{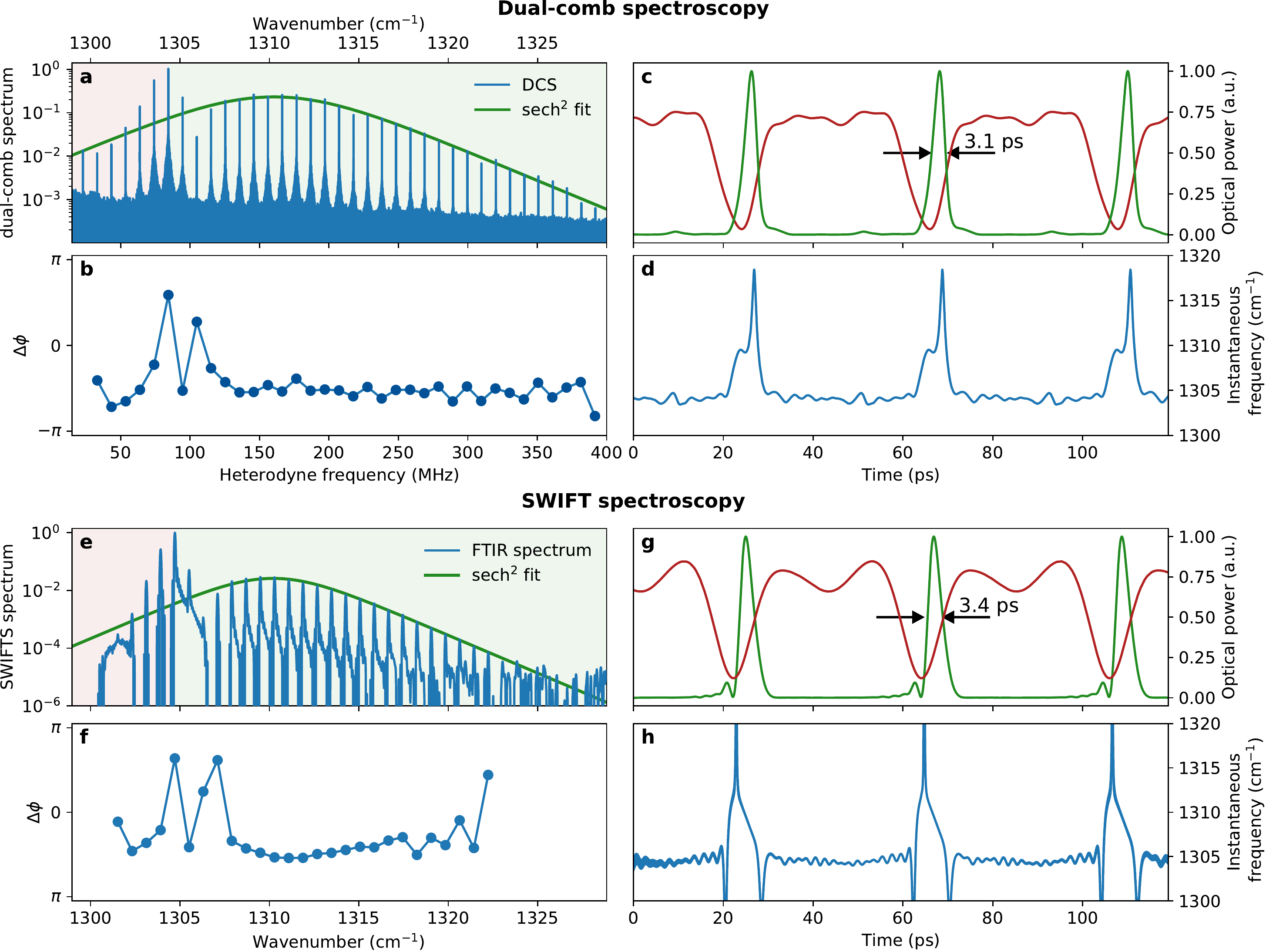}
	\caption{\textbf{Spectral phases and waveform of the QCL solitons.} \textbf{a}, Dual-comb spectrum at 966 mA. The red and green shaded region denote normal and anomalous dispersion, respectively. \textbf{b}, Intermodal difference phases $\Delta \phi$ of the ring QCL. The sech$^2$-shaped part of the spectrum shows in-phase synchronization, as expected for a soliton.  \textbf{c}, Reconstructed waveform of the sech$^2$ shaped part of the spectrum ($>1306$~cm$^{-1}$, green curve) and the dispersive wave ($<1306$~cm$^{-1}$, red curve). \textbf{d}, Instantaneous wavenumber of the waveform in \textbf{c}. \textbf{e}, SWIFTS spectrum at 966 mA. \textbf{f}, $\Delta \phi$ extracted from the SWIFTS spectrum. \textbf{g}, Reconstructed waveform using SWIFTS (green curve: $>1306$~cm$^{-1}$, red curve: $<1306$~cm$^{-1}$). \textbf{h}, Instantaneous wavenumber retrieved by SWIFTS.}
	\label{fig:fig3}
\end{figure*}

A key advantage of dual-comb spectroscopy is the heterodyne gain provided by the powerful reference comb ($>$100~mW average power), enabling extremely sensitive detection. After coherent averaging\cite{sterczewski2019computational}, the dual-comb spectrum (Fig. \ref{fig:fig3}a) reveals almost 40 teeth of the ring QCL spectrum, whose linewidth is limited by the acquisition time over the entire spectral bandwidth. This result underlines the high temporal coherence of the solitons emitted by the ring QCL, which is also evident from the narrow beatnote in Fig. \ref{fig:fig2}. In close agreement with SWIFTS (Fig. \ref{fig:fig3}e), the dual-comb spectrum consists of a group of strong modes around the zero dispersion point and a sech$^2$-shaped lobe spanning over 20~cm$^{-1}$ in the anomalous dispersion regime. Most importantly, the intermodal phase differences ($\Delta \phi$, Fig. \ref{fig:fig3}b) of the sech$^2$-shaped part of the spectrum are synchronized in-phase, which is a clear evidence for the formation of optical solitons. Conversely, the strongest mode around 1304~cm$^{-1}$ is out of phase by $\pi$ compared to its neighboring modes and the sech$^2$ lobe. Therefore, the strong modes around the zero dispersion point form a dispersive wave, which closely resembles an ideal frequency modulated wave with a low modulation index. This observation is confirmed by the reconstructed waveform shown in Fig. \ref{fig:fig3}c. The peak of the spectrum around the zero dispersion point gives rise to a quasi-continuous wave, interrupted by a short period of almost zero intensity once per roundtrip. During this time window, the modes of the sech$^2$-shaped lobe interfere constructively to form a bright soliton with approximately 3.1~ps full width at half maximum. The instantaneous frequency (Fig. \ref{fig:fig3}d) further validates that the soliton arises from the sech$^2$-shaped lobe of the spectrum.
Both the spectrum and the intermodal phase differences measured using SWIFTS (Figs. \ref{fig:fig3}e,f) show excellent agreement with the dual-comb results. The reconstructed SWIFTS waveform confirms the formation of 3.4~ps bright solitons superimposed on a dispersive wave  (Figs. \ref{fig:fig3}g,h). The slight difference of the pulse width in SWIFTS is attributed to the superior dynamic range of the dual-comb spectroscopy measurement.

\section*{Discussion}

The dynamics of the field inside the ring QCL are governed by the Maxwell-Bloch equations. Recent work~\cite{columbo2021unifying} showed that these equations can be mapped, within certain limits, onto the Lugiato-Lefever equation~\cite{lugiato1987spatial}, which is widely used to model dissipative Kerr solitons in passive microresonators. As a result, we can corroborate the experimental observations using the well-known criterion of the threshold intracavity power
\begin{equation}
    P^{\text{th}}_{\text{cav}} = \frac{\kappa n_0^3 A_{eff}}{2 \omega_0 n_2} \label{eq:threshold}
\end{equation}
for parametric oscillation\cite{herr2014temporal}. Here, $A_{eff} = 29$ \textmu m$^2$ and $n_0=3.2$ are the effective area and refractive index of the waveguide mode, respectively, and $n_2$ is the Kerr coefficient. $\kappa$ represents the maximum cavity loss rate which needs to be overcome by the parametric gain during soliton operation, and is assumed to be close to the waveguide loss rate due to gain saturation during the pulse. The value of $n_2$ is obtained from simulations using a non-equilibrium Green's function formalism\cite{wacker_nonequilibrium_2013} and $n_2 = 3\cdot10^{-14}$ to  $4\cdot10^{-13}$ m$^2$/W from threshold to roll-over. 
The waveguide losses were measured on a Fabry-P\'erot QCL with equal cavity dimensions and were found to be $\sim$3.5 cm$^{-1}$, equivalent to $\kappa = 33$ ns$^{-1}$ and a Q-factor of 7480. The simulations and measurements used to obtain the values above are described in detail in the Supplementary Information. Using these numbers, Eq.~\eqref{eq:threshold} predicts an upper bound for the threshold intracaity power for parametric oscillation of 500~mW around the observed soliton threshold. The actual intracavity power around the symmetry breaking point, where comb formation is first observed, can be inferred from the photon driven current (see Supplementary Information) and is 190~mW, consistent with the simulation results.

The QCL still operates in the bi-directional regime in a small range around the symmetry breaking step in Fig. \ref{fig:fig1}b. Nevertheless, frequency comb operation with a narrow beatnote is observed Fig. \ref{fig:fig1}c. However, neither the spectrum nor the spectral phases of this state (see Supplementary Information) are compatible with a fundamental soliton. This state is equivalent to the frequency combs induced by phase turbulence reported in earlier work\cite{piccardo2020frequency}. Hence, uni-directional lasing is a crucial condition for dissipative Kerr solitons in ring QCLs.


\section*{Conclusion}
The presented results prove unambiguously that the parametric gain originating from the giant Kerr nonlinearity of QCLs is sufficiently large to generate dissipative Kerr solitons. This giant nonlinearity is a property common to most semiconductor lasers emitting from visible wavelengths over the near- and mid-IR regions all the way to the Terahertz domain. Our approach is therefore promising for microcombs in spectral regions where generating dissipative Kerr solitons in passive microresonators is challenging, such as the visible and mid-IR domain. The latter is particularly important for optical sensing and chemical spectroscopy due to the presence of characteristic absorption lines of many molecules - the so-called molecular fingerprint. Although mid-IR Kerr combs were observed in MgF$_2$\cite{wang2013mid} and silicon microresonators equipped with a p-i-n junction to reduce free carriers\cite{griffith2015silicon,yu2018silicon}, the increased material absorption has so far rendered mid-IR dissipative Kerr solitons an elusive goal. The availability of stable soliton microcombs emitting in the mid-IR region is a key step towards integrated and battery driven spectrometers. We predict that also semiconductor lasers emitting in the near-IR range offer an attractive platform for soliton generation provided that they exhibit both the required low level of backscattering and negative dispersion. Structures based on quantum well and quantum dot lasers\cite{rafailov2007mode,auth2019passively} are particularly promising, since they allow to pump the soliton with a short and intense pulse\cite{obrzud2017temporal}.

\section*{Methods}
\subsection*{Ring QCL fabrication}
The active region of the QCLs is based on a bound-to-continuum design using strain-\\compensated InGaAs\textbackslash AlInAs~\cite{wolf2017quantum} and is grown by molecular beam epitaxy. The 10 $\mu$m wide waveguides are defined by wet-chemical etching based on a HBr solution. Due to the isotropic etch profile of HBr, this step enables extremely smooth sidewalls of the waveguide, which are essential for reducing backscattering and thus for uni-directional lasing. Subsequently, semi-insulating Fe:InP is regrown on the sidewalls by metal organic vapor phase epitaxy. This regrowth process significantly reduces the refractive index contrast compared to normal ridge waveguides, thus further suppressing the backscattering. Finally, Ohmic contacts are evaporated on both electrodes of the QCL and the chip is cleaved and mounted junction-down on a AlN heat spreader.

\subsection*{Dual-comb spectroscopy}
 The light emitted by the ring QCL is combined with a reference Fabry-P\'erot QCL comb on a Vigo PV-4TE-10.6 Mercury-Cadmium-Tellurium (MCT) detector with 1 GHz electrical bandwidth. The power of the reference comb is attenuated to approximately 10 mW to avoid detector saturation. The dual-comb beating is sampled at 2.5 GS/s with a LeCroy HDO6104 oscilloscope. The detailed data treatment is explained in the Supplementary Information. 
 
\subsection*{Shifted wave interference Fourier transform spectroscopy}
The QCL emission is collimated using an immersion lens with 3 mm focal length and shined through an FTIR spectrometer. At the exit of the FTIR, a high-speed QWIP detects the fundamental beatnote of the ring QCL. This beatnote is down-mixed to 20~MHz and its in-phase and quadrature component are recorded as function of the FTIR delay using an UHFLI lock-in amplifier. The Fourier transform of the quadrature interferograms yields the spectral phases, as elaborated in the Supplementary Information.

\section*{Data Availability Statement}
The data in this study is available from the corresponding author upon reasonable request.


\small
\bibliographystyle{naturemag}
\bibliography{Bocitation}

\section*{Acknowledgements} 
This work was supported by the Swiss National Science Foundation and Innosuisse in the scope of the CombTrace (No. 20B2-1\_176584/1) project and the Qombs Project funded by the European Union’s
Horizon 2020 research and innovation program under
Grant Agreement (No. 820419). The authors thank Zhixin Wang for support with the high-reflection coatings of the reference Fabry-P\'erot QCLs and Mathieu Bertrand for initial help with device characterization.

\section*{Author contributions statement}
B.M. initiated the project. B.M. processed the devices and performed initial characterization.  M.S. and B.M. performed the SWIFTS measurement. J.H. performed the dual-comb characterization. M.F. provided the NEGF simulations. M.B. grew the QCL wafer. J.H., B.M., M.F and J.F. wrote the manuscript and the Supplementary Information. J.F. supervised the project. All authors contributed to analyzing and interpreting the results. 

\section*{Competing interests} 
The authors declare no competing interests.

\section*{Correspondence} 
*Correspondence should be addressed to B.M.\\(email: bmeng@phys.ethz.ch) or J. F. (email: jerome.faist@phys.ethz.ch).

\end{document}